\pdfoutput=1
\documentclass[%
 reprint,
 amsmath,amssymb,
 aps,
 onecolumn,
prf,
]{revtex4-2}

\usepackage{graphicx}
\usepackage{dcolumn}
\usepackage{bm}
\usepackage{amsmath}
\usepackage{float}
\usepackage{caption}
\usepackage{subcaption}
\usepackage{mathrsfs}

\usepackage{psfrag}
\usepackage{pstool}
\usepackage{pictex}
\begin{document}

%
%
%
%
%
%
%
%
\title{Model-Based Estimation of Vortex Shedding in Unsteady
Cylinder Wakes}
%
%
%
\author{Jiwen Gong, Jason P. Monty and Simon J. Illingworth}
\affiliation{The University of Melbourne, Victoria 3010, Australia}

\begin{abstract}
This paper considers single-sensor estimation of vortex shedding in cylinder wakes at $Re=100$ in simulations and at $Re=1036$ in experiments. A model based on harmonic decomposition is developed to capture the periodic dynamics of vortex shedding. Two model-based methods are proposed to estimate time-resolved flow fields. First, Linear Estimation (LE) which implements a Kalman Filter to estimate the flow. Second, Linear-Trigonometric Estimation (LTE), which utilizes the same Kalman Filter together with a nonlinear relationship between harmonics of the vortex shedding frequency. LTE shows good performance and outperforms LE regarding the reconstruction of vortex shedding. Physically this suggests that, at the Reynolds numbers considered, the higher harmonic motions in the cylinder wake are slave to the fundamental frequency.
\end{abstract}
\maketitle
%
%
\section{Introduction}
The ability to accurately estimate flow fields is needed for a diverse range of fluid mechanics applications. These include flow prediction, drag reduction, and separation control. Generally, flow estimation aims to reconstruct the spatial and temporal evolution of the flow of interest, using limited time-resolved measurements. Flows are commonly estimated using statistical tools or dynamic models. Perhaps the most widely used statistical tool in flow estimation is Linear Stochastic Estimation (LSE) \cite{adrian1977role, guezennec1989stochastic}. The popularity of LSE has led to a number of modifications, including POD-LSE \cite{bonnet1994stochastic}, Spectral LSE \cite{tinney2006spectral, baars2014proper} and Multi-Time-Delay LSE \cite{durgesh2010multi}.

Alternatively one can estimate flows using dynamic models. Model-based estimation requires a dynamic model to describe the evolution of the flow with time, and an estimator to estimate the flow using limited measurements. Generally, system modelling can be either physics-driven or data-driven. Physics-driven approaches derive models based on the Navier-Stokes equations, including the Galerkin-POD method \cite{rowley2004model} and a recently developed resolvent-based method \cite{mckeon2010critical, gomez2016reduced, illingworth2018estimating}. Alternatively, models can be built via data-driven approaches, including Dynamic Mode Decomposition (DMD) \cite{schmid2010dynamic}, Artificial Neural Networks (ANN) \cite{siegel2008low}, Sparse Identification of Nonlinear Dynamics (SINDy) \cite{loiseau2018sparse}, and the Eigensystem Realization Algorithm (ERA) \cite{yao2017feedback, illingworth2016model}.

These various estimation methods have been applied across a broad range of flows, including cavity flows \cite{rowley2005model, murray2007modified, gomez2016reduced}; wall-bounded flows \cite{illingworth2018estimating}; jets \cite{beneddine2017unsteady}; and the flow over a flat plate \cite{tu2013integration}, an airfoil \cite{leroux2018time} and a cylinder \cite{gomez2016estimation, loiseau2018sparse}. This paper focuses particularly on the estimation problem in cylinder wakes, for which the flow first becomes unsteady (i.e. vortex shedding appears) at low Reynolds numbers ($Re>49$) \cite{williamson1996vortex}. Previous work has mostly focused on the estimation of the cylinder wake in the laminar vortex shedding regime ($49<Re<188.5$ \cite{barkley1996three}), and has done so mostly in simulations \cite{gomez2016estimation, loiseau2018sparse}. The aim of this work is to develop estimation methods that i) can be applied in both simulations and experiments; ii) can be applied at higher Reynolds numbers where turbulent structures also appear.

The paper is organized as follows. Section \ref{s2} introduces the computational cylinder flow at $Re= 100$ and the experimental cylinder flow at $Re= 1036$. The harmonic-based system modelling procedure is introduced in Section \ref{s3}. The two estimation methods, Linear Estimation and Linear-Trigonometric Estimation, are first introduced in Section \ref{s4}, and then applied to the $Re= 100$ case and the $Re= 1036$ case in Section \ref{s5}. Conclusions are drawn in Section \ref{s6}.


\section{Computational and Experimental Setup}\label{s2}
This section introduces the two setups considered in this paper: the unsteady cylinder wake at $Re= 100$ in simulations and the unsteady cylinder wake at $Re= 1036$ in an experiment.
\subsection{Computational Setup}
The cylinder wake at a Reynolds number of $Re= 100$ is solved using Direct Numerical Simulation (DNS) in two dimensions using FEniCS \cite{logg2012automated}. The simulation follows the numerical scheme used in \cite{jin2018resolvent}. The Reynolds number is defined as $Re=\frac{UD}{\nu}$, where $U$ is the freestream velocity, $D$ is the cylinder diameter and $\nu$ is the kinematic viscosity. The spatial discretization is performed using a Taylor-Hood finite element scheme. A first-order difference scheme is used for time integration. The final mesh contains $5.46\times 10^4$ triangles, with a smallest radial increment of 0.01 near the cylinder. Once the flow reaches the limit-cycle state, 5400 instantaneous velocity fields are recorded (with a time step between them of $\Delta t=0.01s$). Together these constitute approximately nine shedding cycles. These are divided into a modelling set (600 time steps) for system modelling and a reference set (4800 time steps) for estimation.

\subsection{Experimental Setup}
\begin{figure}[H]
\centering
\includegraphics[width=.8\textwidth]{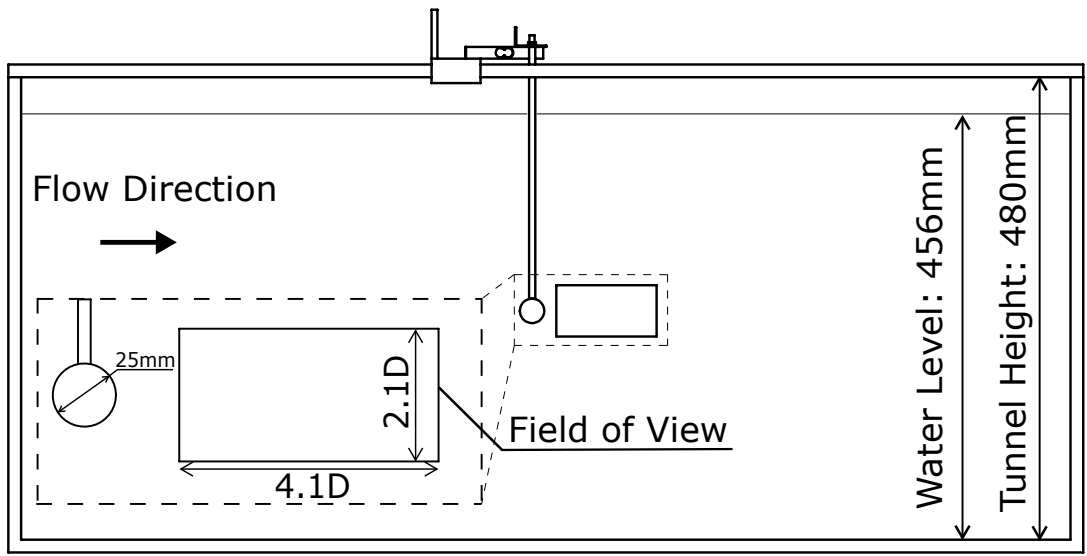}
\caption{A schematic showing the experimental setup and the field of view for PIV.}
\label{fig:tunnel}
\end{figure}

The cylinder wake at a Reynolds number of $Re= 1036$ is conducted in the water tunnel at the University of Melbourne. A schematic of the experimental setup is shown in figure \ref{fig:tunnel}. The freestream velocity is $U_f=41.5$ mm/s. The tunnel has an open top, from which the cylinder is suspended in water. The test cylinder is a sealed aluminum pipe with outer diameter $D = 25$ mm and a thickness of $3$ mm. The total length of the cylinder is $495$ mm giving an aspect ratio of $19.8$.  

The velocity fields are obtained using a Time-Resolved Particle Image Velocimetry (TR-PIV) setup. Flow snapshots are recorded at 50Hz for 6001 time steps, with a spatial resolution of $1920\times1080$ pixels. This frame rate ensures that at least 50 snapshots are available for one vortex shedding period. A 60-watt continuous $532$nm laser source is placed at the bottom of the tunnel, providing a laser sheet with $3$ mm thickness perpendicular to the cylinder axis. The tracer particles used are silver-coated hollow glass spheres ($10$ $\mu$m in diameter) from DANTEC DYNAMICS. The field of interest (shown in Figure \ref{fig:tunnel}) covers the vortex formation region and a downstream region containing approximately two vortices simultaneously. An image background subtraction method \cite{adrian2011particle} is used to remove the background patterns consistently appearing in the raw images. Abnormal measurements are filtered out using a $3\times 3$ median number filter. To obtain the velocity fields, a multi-grid interrogation method is used. In the first stage, the window size is $64\times 64$ with an overlap of $0.5$. In the second stage, a smaller window size of $32\times 32$ pixels is used with the same overlap. Finally, the 6000-time-step velocity fields are divided into two equally sized sets for system modelling and estimation.  \\

\section{Harmonic-Based System Modelling}\label{s3}
\subsection{Harmonics in cylinder flows}
\label{Harmonics}
\begin{figure}[h]
\centering
\includegraphics[width=\textwidth]{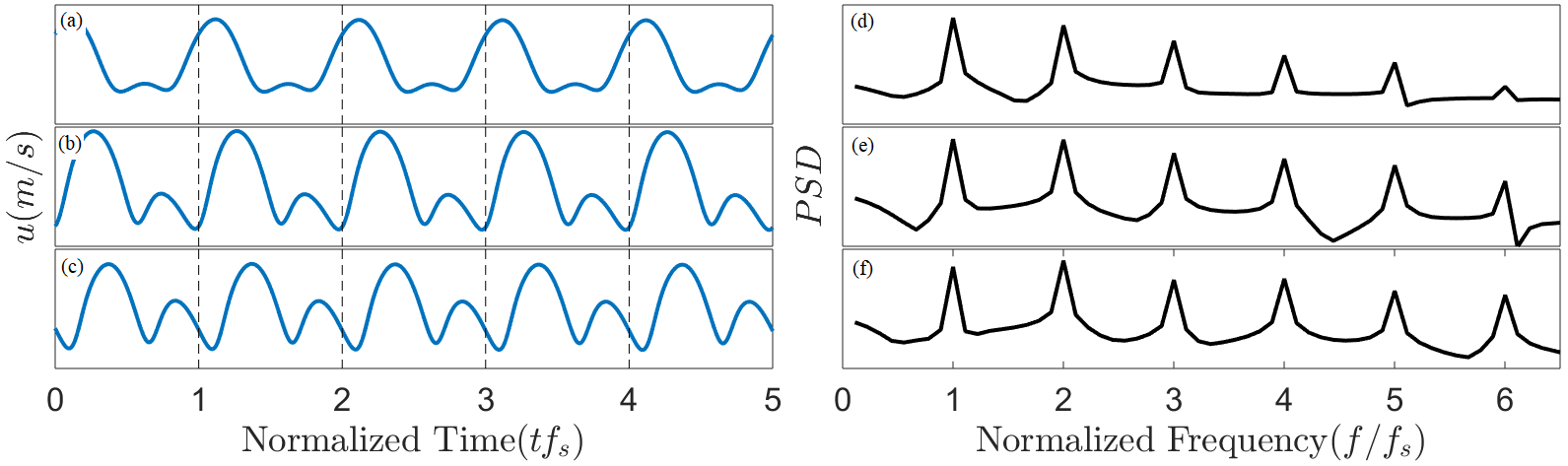}
\caption{The time series and the spectrum of the streamwise velocity fluctuation at $\boldsymbol{x} = (D, 0)$ (a, d), $\boldsymbol{x} = (2D, 0)$ (b, e) and $\boldsymbol{x} = (2.5D, 0)$ (c, f). Time has been normalized by the vortex shedding frequency ($t^*=t f_s$). Frequency has been normalized by the vortex shedding frequency ($f^*=\frac{f}{f_s}$).}
\label{fig:Re100_u}
\end{figure}

Periodicity is a key feature of vortex shedding in cylinder flows \cite{williamson1996vortex}. The normalized vortex shedding frequency, known as the Strouhal Number, has been widely used to characterize the periodic features of vortex shedding \cite{williamson1988existence, prasad1997three, norberg2003fluctuating}. However, the vortex shedding frequency is not the whole story, particularly when it comes to flow reconstruction and modelling. Figure \ref{fig:Re100_u}(a-c) show the temporal signals of the streamwise velocity fluctuations at different locations in the cylinder wake at $Re=100$. The velocity signals all repeat themselves at the same frequency (vortex shedding frequency), but they have different shapes. These signals are therefore periodic but not sinusoidal. Therefore, from a flow-reconstruction point of view, knowing only the dominant frequency might not be sufficient. 

The key to characterizing the shape of a non-sinusoidal periodic signal is its higher harmonics. Figure \ref{fig:Re100_u}(d-f) show the velocity signals in the frequency domain. Clear peaks at the harmonics of the vortex shedding frequency can be found. Intuitively, the fundamental frequency (vortex shedding frequency) determines the period of a velocity signal, and its harmonics determine the signal shape within one period. Therefore, the vortex shedding frequency and its harmonics all need to be considered when reconstructing the vortex shedding patterns. 

Gathering information about the higher harmonics is difficult because velocity signals are dominated by the vortex shedding frequency. Measurement noise in experimental cases makes it even more difficult to extract higher harmonics. Therefore in this work, a phase averaging method, based on Proper Orthogonal Decomposition (POD), is used to filter out the energy of sensor noise and any aperiodic structures. POD captures the most energetic structures in the flow, which makes it suitable for identifying the vortex shedding phase in cylinder flows \cite{van2005phase}. The POD-based method proceeds in two parts. First, we compute the phases (in time) of the two most energetic POD modes. Second, we phase-average the flow using these phases. The phase-averaged flow is then used to build the low-order model of the flow in the following subsections.

\subsection{Harmonic Decomposition}
\label{Sec:HD}
The strong harmonic peaks suggest the possibility of finding a low-order representation of the fluid system based on these harmonics. Here we use Harmonic Decomposition, a decomposition method used in Harmonic Balance methods\cite{hall2002computation}, to decompose the flow based on the vortex shedding frequency and its harmonics. Using the same notation introduced in \cite{liu2006comparison}, the phase-averaged velocity field is decomposed into two parts: i) a time-independent mean flow, and ii) a sum of harmonic modes. This decomposition of the velocity field gives: 
\begin{eqnarray}
u_i &=& \Bar{u}_i + \sum_{n=1}^{N}[c_{i,2n-1} \cos(n\omega t) +c_{i,2n} \sin(n\omega t)],
\label{eq:hd1}
\end{eqnarray}
where $\omega$ is the known fundamental harmonic (i.e. vortex shedding frequency). The terms $c_{i,2n-1}$ and $c_{i,2n}$ contain the time-independent coefficients of the $nth$ harmonic at grid point $i$; and $N$ is the number of harmonics retained. To further extend \eqref{eq:hd1} to the entire domain, one can represent the evolution of the velocity fluctuations of the entire field in terms of a finite number of harmonics and their corresponding modeshapes $\Phi$, given as: 
\begin{eqnarray}
U &=&\Bar{U} + \Tilde{U},\\
\Tilde{U}&=&\Phi X,
\label{eq:hd2}
\end{eqnarray}
where $U$ contains the two-dimensional velocity fields as column vectors. The term $\Bar{U}$ corresponds to the mean velocity field, and $\Tilde{U}$ corresponds to the fluctuations about the mean. The term $\Phi$ represents row vectors of modeshapes containing the coefficients $c$. The term $X$ contains all harmonics. These terms are given as:
\begin{eqnarray}
\begin{aligned}[t]
U &= \begin{bmatrix} 
    u_1(t_1) & \dots  & u_1(t_M)\\
    \vdots   & \ddots & \vdots\\
    u_I(t_1) & \dots  & u_I(t_M)\\
    v_1(t_1) & \dots  & v_1(t_M)\\
    \vdots   & \ddots & \vdots\\
    v_I(t_1) & \dots  & v_I(t_M)\\
\end{bmatrix}, \nonumber\\
\Phi &= \begin{bmatrix} 
    c_{1,1} & \dots  & c_{1,(2N)}\\
    \vdots       & \ddots & \vdots\\
    c_{2I,1} & \dots  & c_{2I,(2N)}\\
\end{bmatrix} \nonumber\\
     &= \begin{bmatrix} \phi_1&\dots&\phi_{2N}\end{bmatrix},\nonumber
\end{aligned}
\qquad \qquad \qquad
\begin{aligned}[t]
\tilde{U} &= \begin{bmatrix} 
    \tilde{u}_1(t_1) & \dots  & \tilde{u}_1(t_M)\\
    \vdots           & \ddots & \vdots\\
    \tilde{u}_I(t_1) & \dots  & \tilde{u}_I(t_M)\\
    \tilde{v}_1(t_1) & \dots  & \tilde{v}_1(t_M)\\
    \vdots           & \ddots & \vdots\\
    \tilde{v}_I(t_1) & \dots  & \tilde{v}_I(t_M)\\
\end{bmatrix},\nonumber\\
X &= \begin{bmatrix} 
    \cos(\omega t_1) & \dots  & \cos(\omega t_M)\\
    \sin(\omega t_1) & \dots  & \sin(\omega t_M)\\
    \vdots    & \ddots & \vdots\\
    \cos(N\omega t_1) & \dots  & \cos(N\omega t_M)\\
    \sin(N\omega t_1) & \dots  & \sin(N\omega t_M)\\
\end{bmatrix} \nonumber\\
  &= \begin{bmatrix} x(t_1)&\dots&x(t_M)\end{bmatrix},\nonumber
\end{aligned}
\end{eqnarray}
where $I$ is the number of grid points and $M$ is the number of time steps. $u_i(t_m)$ and $v_i(t_m)$ represent the streamwise and transverse velocity at time $t=t_m$, respectively. Each harmonic has two modeshapes ($\phi_{2n-1}$ and $\phi_{2n}$), corresponding to $\cos(n\omega t)$ and $\sin(n\omega t)$, respectively. $x(t_m)$ contains the time dynamics of all modes at time $t=t_m$.

A more common method to find modeshapes based on frequencies is Dynamic Mode Decomposition (DMD) \cite{schmid2010dynamic}.
DMD finds active frequencies in the flow and generates corresponding modeshapes.
DMD presents two challenges for the estimation problem of interest in this work.
First, DMD can give rise to multiple modes with similar temporal frequencies---particularly when the flow exhibits turbulent features. Second, it can be challenging to rank DMD modes based on their relative importance. For these reasons we choose to use harmonic decomposition in this work, since only one mode is obtained for each harmonic, and ranking the modes is straightforward.


To summarize, harmonic decomposition decomposes the evolution of the velocity fields $U(x,y,t)$ into time-independent modeshapes $\Phi(x,y)$ and a time-dependent dynamic term $X(t)$. It significantly reduces the number of time-dependent variables from twice the number of grid points to twice the number of harmonics chosen. Less time-dependent variables allow a reduced-order model to be formed in a data-driven way in the following subsection.

\subsection{Data-Driven Model Building}
The model building procedure consists of two parts: i) finding modeshapes; and ii) forming a dynamic model.

\textbf{Finding modeshapes:} A least-square-error method is used to fit the harmonic modeshapes $\Phi$. The method requires knowledge of the phase-averaged velocity fields and the vortex shedding frequency, which can come either from simulations or from experimental measurements. The method gives
\begin{eqnarray}
\hat{\Phi} &=& \Tilde{U}_{pa} X^T(XX^T)^{-1},
    \label{eq:sf}
\end{eqnarray}
where $\hat{\Phi}$ are the fitted modeshapes and $\Tilde{U}_{pa}$ are the velocity fluctuations of the phase-averaged flow. Direct measurements of the velocity field $\Tilde{U}$ may be used to replace $\Tilde{U}_{pa}$ in \eqref{eq:sf} if measurement noise is negligible and any turbulence is weak.

\textbf{Forming a dynamic model:} The  term $X(t)$, which represents sinusoidal signals with multiple temporal frequencies, is modelled as a marginally stable linear oscillator with multiple harmonic frequencies. The model is of the form
\begin{eqnarray}
\dot{x}(t) &=& A x(t),
\end{eqnarray}
where $x$ is the state of the model. The state matrix $A$ is
\begin{equation}
A = \begin{bmatrix}
    A_{\omega_1} & \cdots & 0\\
    \vdots & \ddots & \vdots\\
    0 & \cdots & A_{\omega_N}
    \end{bmatrix}, \hspace{5mm} 
A_{\omega_n} = \begin{bmatrix}
    0 & -n\omega\\
    n\omega & 0
    \end{bmatrix},
\label{eq:A_omega}
\end{equation}
where $\omega_n$ indicates the $n$th harmonic, and $N$ is the number of harmonics considered in the model. 

\textbf{State-space model:} Combining the fitted modeshapes $\hat{\Phi}$ and the linear model of the time dynamics $x$, a state-space model of the velocity fluctuation is formed,
\begin{eqnarray}
\dot{x}(t) &=& A x(t),\nonumber\\
\Tilde{U}(t) &=& \hat{\Phi} x(t),
\label{eq:output}
\end{eqnarray}
where $\Tilde{U}(t)$ represents one column of $\Tilde{U}$ at time $t$. The modeshape matrix $\hat{\Phi}$ is regarded as an output matrix in the model. The state-space model is then used to design an estimator in the following section. 

\section{Estimation Methods}\label{s4}
In this section we introduce two model-based estimation methods: Linear Estimation (LE) and Linear-Trigonometric Estimation (LTE). The model described in Section \ref{s3} is used for both methods.

\subsection{Linear Estimation}
The goal of Linear Estimation is to estimate the entire velocity field based on limited velocity sensor measurements (one sensor in the current case). LE includes two steps: i) augment the linear model to include system disturbances, sensor measurement and measurement noise and ii) design a dynamic estimator based on the augmented linear model. Once the dynamic estimator is obtained, the velocity fields can be estimated by feeding the sensor measurement to the dynamic estimator.

In the first step, we augment the model described in Section \ref{s3} (Equation (\ref{eq:output})). The augmented model now takes the form:
\begin{eqnarray}
\dot{x}(t) &=& A x(t) + w(t),\nonumber\\
\Tilde{U}(t) &=& \hat{\Phi} x(t),\nonumber\\
u_m(t) &=& C_m x(t)+s(t),
\label{eq:model}
\end{eqnarray}
where $u_m(t)$ is the single sensor measurement, and $C_m$ is the corresponding output matrix, which contains a certain row of the modeshape matrix $\hat{\Phi}$. This row contains the values of the modeshapes for all the harmonics at the sensor location. The term $w(t)$ represents disturbances to the system, and $s(t)$ represents measurement noise.

In the second step, we use the augmented model (Equation (\ref{eq:model})) to design a Kalman Filter \cite{kalman1960new}, which is an optimal linear dynamic estimator in the sense of minimizing the amplification of the estimation error from unknown disturbances $w(t)$ and measurement noise $n(t)$. The Kalman Filter uses  knowledge of the sensor measurement $u_m(t)$ to estimate the full state and then the entire velocity field:
\begin{eqnarray}
\dot{\hat{x}}(t) &=& A \hat{x}(t) + L(\hat{u}_m(t)-u_m(t)),\nonumber\\
\hat{\Tilde{U}}(t) &=& \hat{\Phi}\hat{x}(t),
\label{eq:kal}
\end{eqnarray}
where $\hat{x}(t)$ is the estimate of the full state $x(t)$; $\hat{\Tilde{U}}(t)$ is the estimate of the velocity field $\Tilde{U}(t)$; and $\hat{u}_m(t)$ is the estimate of the sensor measurement. The Kalman gain $L$ is obtained by solving a Riccati equation, which ensures that the error $e(t)=\hat{x}(t)-x(t)$ converges to zero. 

Essentially, LE builds linear transfer functions from the sensor measurement to both the fundamental harmonic and higher harmonics. However, it has been pointed out in \cite{rosenberg2019role} that linear mechanisms are not dominant at higher harmonics in cylinder flows. Therefore, two questions may be asked. First, can LE estimate the higher harmonics, for which linear mechanisms are not dominant? We will discuss this in more detail in Section \ref{s5}. Second, can we estimate the higher harmonics nonlinearly? To answer the second question, we will introduce Linear-Trigonometric Estimation in the following.  

\subsection{Linear-Trigonometric Estimation}
\begin{figure}[h]
\centering
\includegraphics[width=.9\textwidth]{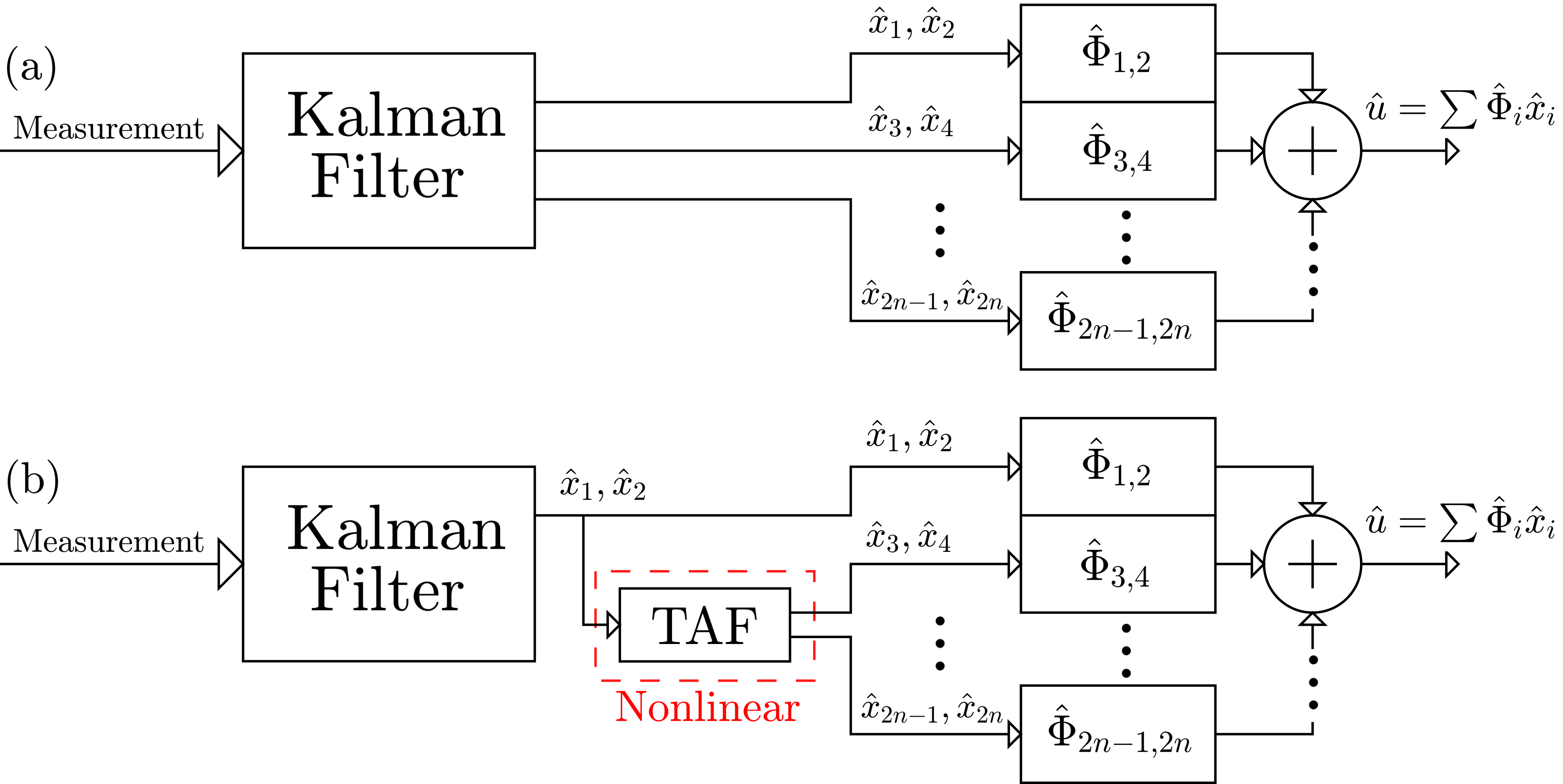}
\caption{Schematic Diagram of (a) LE and (b) LTE.}
\label{fig:LE_LTE}
\end{figure}

The idea behind Linear-Trigonometric Estimation is to include the nonlinear trigonometric relationships among harmonics into Linear Estimation. LTE uses the same method as LE to linearly estimate the dynamics of the fundamental harmonic. The difference is that LTE then nonlinearly estimates the time dynamics of the higher harmonics from the estimate of the fundamental harmonic. We can do this because the term $x(t)$ only contains disturbed sinusoidal signals at different harmonics, and all the information about their phases and amplitudes are averaged and stored in modeshapes $\Phi$. By introducing the nonlinear operator $\mathscr{N}$, the output equation (Equation (\ref{eq:kal})) is augmented as,
\begin{eqnarray}
\hat{\Tilde{U}}(t) &=& \hat{\Phi}\begin{bmatrix} \hat{x}_{1}(t) & \hat{x}_{2}(t) & \hdots & \hat{x}_{2N-1}(t) & \hat{x}_{2N}(t)\end{bmatrix}^T,\nonumber\\ 
&=& \hat{\Phi}\begin{bmatrix} \hat{x}_{1}(t) & \hat{x}_{2}(t) & \mathscr{N}(\begin{bmatrix} \hat{x}_{1}(t) & \hat{x}_{2}(t)\end{bmatrix}^T)\end{bmatrix}^T,
\label{eq:IE}
\end{eqnarray}
where $\hat{x}_{1}(t)$ and $\hat{x}_{2}(t)$ are the estimated states containing the cosine and sine of the fundamental harmonic respectively; and $N$ is the number of harmonics retained for estimation. The term $\mathscr{N}(\begin{bmatrix} \hat{x}_1(t) & \hat{x}_{2}(t)\end{bmatrix}^T)$ contains the cosines and sines of the higher harmonics, where $\mathscr{N}$ is a recursive nonlinear operator based on Trigonometric Addition Formulas (TAF), given as,

\begin{eqnarray}
\begin{bmatrix} x_{2n-1}\\x_{2n}\end{bmatrix} &=& \begin{bmatrix} \cos(n\omega t)\\\sin(n\omega t)\end{bmatrix},   \nonumber\\
&=& \begin{bmatrix} \cos((n-1)\omega t)\cos(\omega t) - \sin((n-1)\omega t)\sin(\omega t) \\\sin((n-1)\omega t)\cos(\omega t) + \cos((n-1)\omega t)\sin(\omega t)\end{bmatrix},\nonumber\\
&=& \begin{bmatrix} x_{2n-3}x_1-x_{2n-2}x_2\\
 x_{2n-2}x_1+x_{2n-3}x_2\end{bmatrix} ,\label{eq:nlr1}
\end{eqnarray}
where $x_{2n-1}$ and $x_{2n}$ are the states containing the time dynamics of the $n$th harmonic, and $x_{2n-1}$ and $x_{2n}$ are the dynamics of the $n$th harmonic. Based on Equation (\ref{eq:nlr1}), any harmonic can be expressed as a nonlinear function of the fundamental harmonic. A schematic is shown in Figure \ref{fig:LE_LTE} to compare the work flows of both LE and LTE. 

\section{Results and Discussions}\label{s5}
The performance of LE and LTE are compared in two cases: the $Re=100$ case in DNS and the $Re=1036$ case in an experiment. 

\subsection{Cylinder flow at $Re=100$}
\label{RE100}
For the $Re=100$ case, the Strouhal Number is $f=0.167$ (obtained from the power spectral density of the velocity), which is consistent with existing literature \cite{norberg2003fluctuating}. The transverse (vertical) velocity is chosen as the input signal because the magnitude of its fluctuation is greater than that of the streamwise velocity. This provides a higher signal-to-noise ratio. The sensor is placed at $x_0 = 2.90D, y_0 = 0.02D$ because this is the location where the perturbation energy is maximized.

\begin{figure}[h]
\centering
\includegraphics[width=\textwidth]{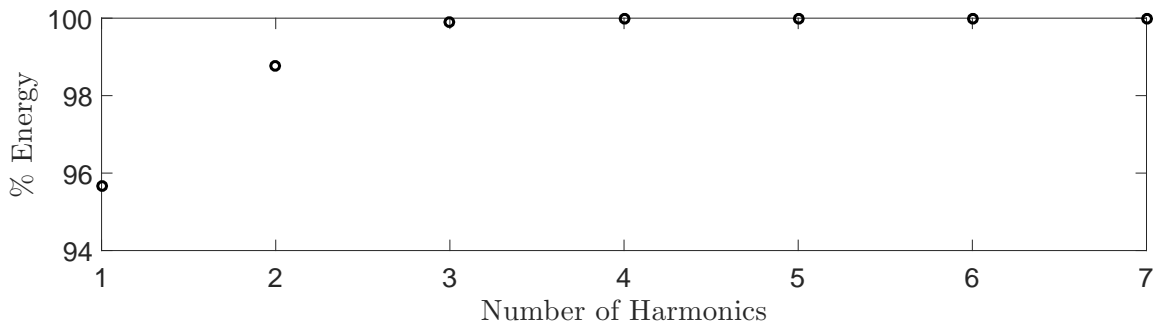}
\caption{Percentage of total energy captured as a function of the number of the harmonics retained.}
\label{fig:Re100_energy}
\end{figure}

First, we examine the reconstruction of the flow fields using the harmonic decomposition (Equation \ref{eq:hd2}). LE estimates the time dynamic term $X(t)$ in Harmonic Decomposition, which means a perfect estimator can only give estimates as good as the reconstruction from Harmonic Decomposition. The reference set of the velocity fields is first decomposed into harmonic modeshapes to obtain the time dynamic term,
\begin{eqnarray}
X(t) = (\hat{\Phi}^T\hat{\Phi})^{-1}\hat{\Phi}^T\tilde{U}.
\label{eq:xt_re100}
\end{eqnarray}
Then the flow field is reconstructed based on Equation (\ref{eq:hd2}). Generally, the more harmonics included in $X(t)$, the higher percentage of energy can be captured. Figure \ref{fig:Re100_energy} shows the percentage of energy captured as more harmonics are included. We observe that higher harmonics contain significantly less energy. Therefore, a third-order harmonic decomposition (first three frequencies) is used in both LE and LTE, capturing 99.9\% of the total perturbation energy.

\begin{figure}[h]
\centering
\includegraphics[width=\textwidth]{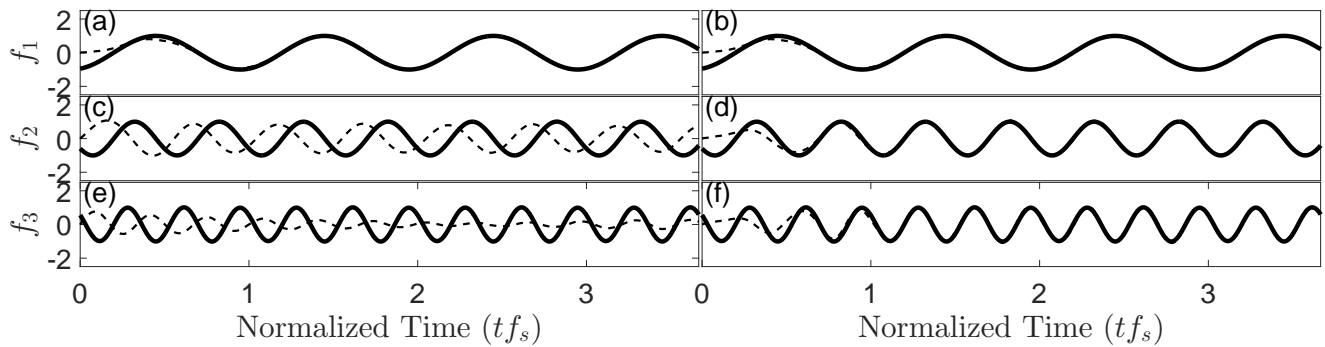}
\caption{Time Coefficients of the Fundamental Harmonic(a,b), the second harmonic(c,d) and the third harmonic(e,f) from DNS(solid lines), Linear Estimation(dashed lines in a,c,e) and Linear-trigonometric Estimation (dashed lines in b,d,f). Time has been normalized by the vortex shedding frequency ($t^*=t f_s$).}
\label{fig:Re100harmonicsDE}
\end{figure}

We now look at the performance of LE and LTE for the first three harmonics, comparing to the harmonic decomposition. Figure \ref{fig:Re100harmonicsDE} compares the time dynamics term $X(t)$ decomposed from the DNS field (solid lines) to the estimate from LE and LTE (dashed line) for the first three harmonics. We observe that the estimates match the DNS curves after a short transient at the fundamental harmonic (Figure \ref{fig:Re100harmonicsDE}(a,b)) for both LE and LTE. This suggests that the dynamic estimator (Kalman filter) works well at the fundamental harmonic using the linear model and the measurements available. 

However, clear discrepancies are observed for higher harmonics in LE (Figure \ref{fig:Re100harmonicsDE}(c,e)). One possible explanation for the poor performance of LE at higher harmonics is: The components of higher harmonics in the measurement signal are significantly weaker than the fundamental harmonic, making it difficult to obtain information about the higher harmonics directly from measurements.

Compared to LE, LTE shows better performance for the higher harmonics (Figure \ref{fig:Re100harmonicsDE}(d,f)). What differs is that LTE estimates the time dynamics of the higher harmonics using the fundamental harmonic and the nonlinear trigonometric relationship among harmonics. Therefore, good estimates from LTE indicate that the time dynamics of the higher harmonics are dependent on the fundamental harmonic. This suggests that the flow features related to the higher harmonics can be regarded as "slave" to the fundamental harmonic. We will discuss this in more detail in Section \ref{s54}.

\begin{figure}[h]
\centering
\includegraphics[width=\textwidth]{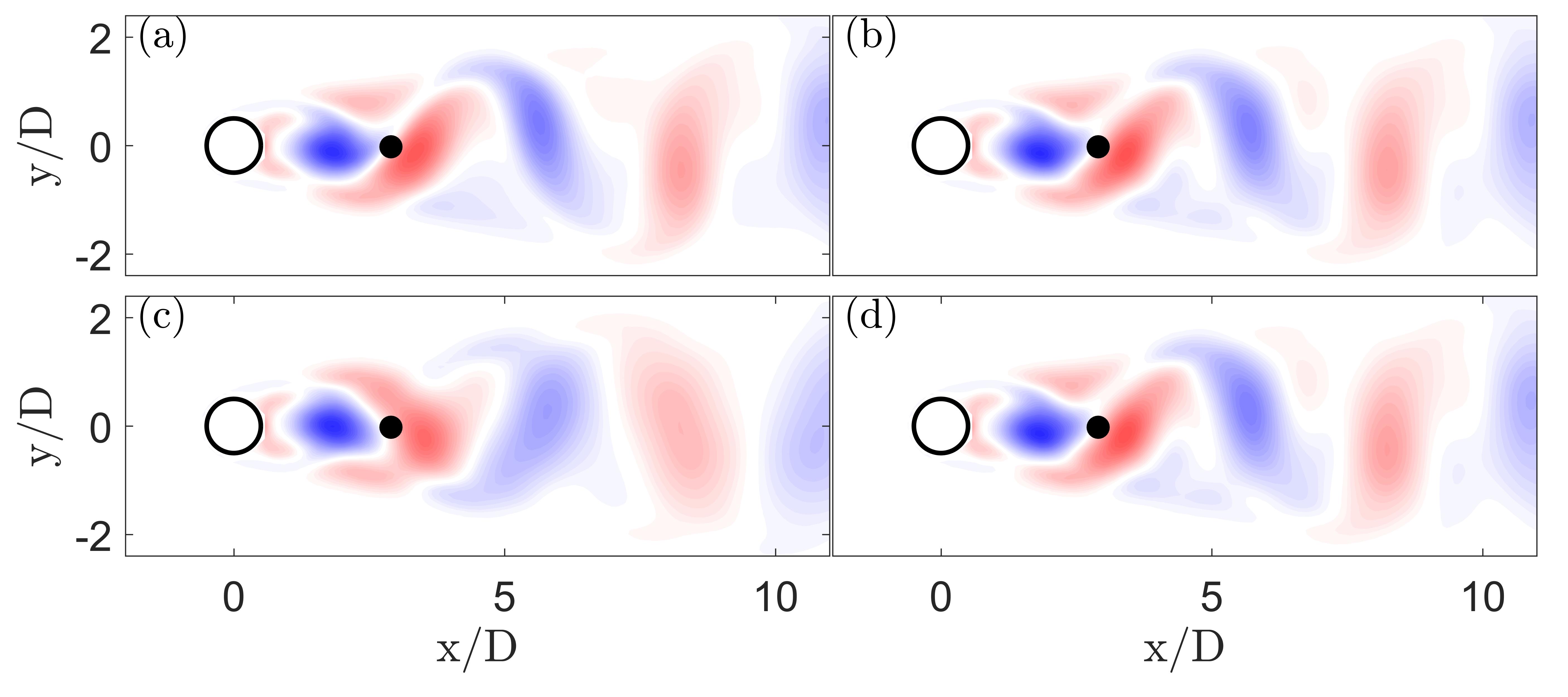}
\caption{The instantaneous vorticity fields from DNS (a), Harmonic Decomposition using first three harmonics (b) and the estimates from Linear Estimation (c) and Linear-Trigonometric Estimation (d). Contour levels for all the plots are $-3$ (blue) to $0$ (white) to $3$ (red).}
\label{fig:Re100_DE_contourf}
\end{figure}

Now we compare the estimates of flow fields from LE and LTE to the harmonic decomposition and the DNS fields. Figure \ref{fig:Re100_DE_contourf} (movies provided) compares instantaneous vorticity fields from DNS, flow reconstruction (3 harmonics), LE and LTE. Overall good agreement can be found between the reconstructed field (Figure \ref{fig:Re100_DE_contourf}(b)) and the DNS field (Figure \ref{fig:Re100_DE_contourf}(a)). This is consistent with the reconstructed flow capturing 99.9\% of the perturbation energy. Comparing the estimates of LE (Figure \ref{fig:Re100_DE_contourf}(c)) and LTE (Figure \ref{fig:Re100_DE_contourf}(d)) to the DNS field (Figure \ref{fig:Re100_DE_contourf}(a)), two observations are made. First, the estimator reasonably estimates the streamwise positions of the vortices and their convective motion for both methods. This is consistent with the observation from Figure \ref{fig:Re100harmonicsDE}(a)(b), where both methods estimate the fundamental harmonic reasonably well. This is because the fundamental harmonic (vortex shedding frequency) determines the period of the flow patterns, which in turn determines the convective velocity of the vortex street. Second, clear discrepancies can be seen in the LE estimates regarding the shapes and the inclination angles of the vortices. This again is consistent with the previous observation of poor estimates of LE at higher harmonics (Figure \ref{fig:Re100harmonicsDE}(c)(e)). This is because, as discussed in Section \ref{Harmonics}, higher harmonics determine the shapes of the velocity fluctuation within one period, which in large shapes the vortices. 

\begin{figure}[h]
\centering
\includegraphics[width=\textwidth]{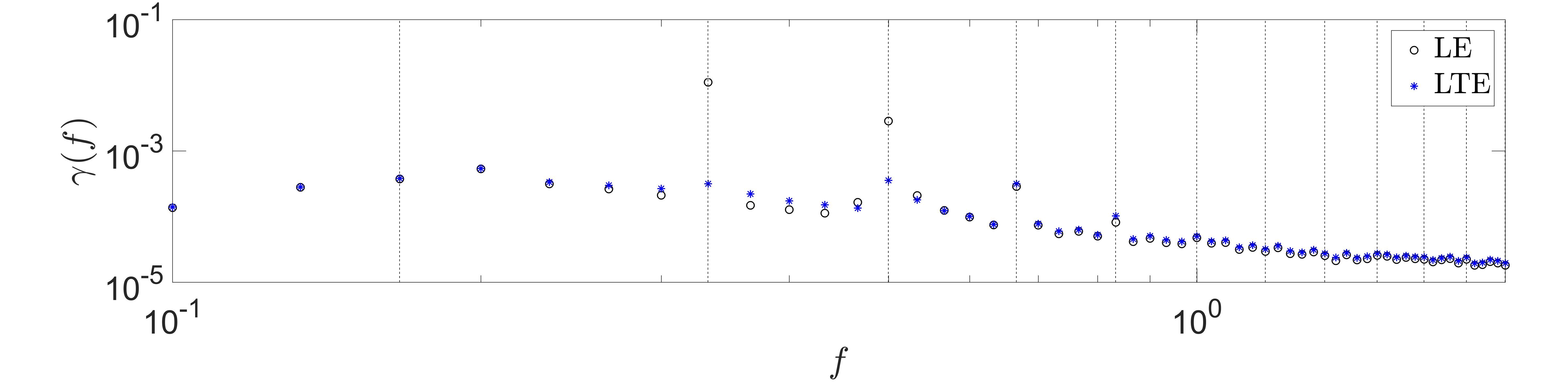}
\caption{Spectrum of the integrated estimation error from LE and LTE. All harmonic frequencies are marked by dashed lines.}
\label{fig:Re100error}
\end{figure}

To quantitatively compare the performance of the two estimation methods, we first introduce a measure of the estimation error $\gamma$:

\begin{eqnarray}
\gamma(f) = \int_y \int_x ||\mathscr{F}\{\hat{\Tilde{U}}(x,y,t)-\Tilde{U}(x,y,t)\}||_2 dxdy,
\end{eqnarray}
where $\mathscr{F}$ represents Fast Fourier Transformation and $||\cdot||_2$ represents the two norm of a complex number. $\hat{\Tilde{U}}$ and $\Tilde{U}$ are the estimate and the true value of the velocity fluctuation respectively. The quantity $\gamma(f)$ is the 2-norm of the spectrum of the estimation error integrated over the entire domain. It represents the distribution of the estimation error (integrated over the entire physical domain) in the frequency domain. Figure \ref{fig:Re100error} compares $\gamma(\omega)$ for LE and LTE. Two observations are made. First, the two methods perform equally well at the fundamental harmonic, which suggests no extra error is introduced (for LTE) at the fundamental harmonic when we non-linearly estimate the time dynamics of the higher harmonics. Second, LTE performs significantly better than LE at the second and the third harmonics.

\subsection{Cylinder flow at $Re=1036$}\label{RE1000}
We now look at the $Re= 1036$ experimental case. The Strouhal Number is measured as 0.21, which is consistent with the literature \cite{norberg2003fluctuating}. Following the same configuration as the $Re=100$ case, we use the PIV measurements at a single location as the sensor measurement, measuring the transverse velocity fluctuation at $x_0 = 2.8D, y_0 = 0$ , where the perturbation energy is the largest. Then we use LE and LTE to estimate the time dynamics of the harmonics, and further reconstruct the entire flow field.
 
Estimation in this case is more challenging than in the Re 100 case for three reasons. First, at this Reynolds number, periodic three-dimensional structures and aperiodic small turbulent structures appear \cite{williamson1996vortex}. This makes it more challenging to obtain the vortex shedding patterns from the raw PIV measurement. Second, disturbances and measurement noise introduce extra uncertainties. Third, the vortex shedding peak observed in the velocity spectrum becomes more broad-band with increasing Reynolds number \cite{norberg1993pressure}, which means the vortex shedding frequency is prone to jitter at this higher Reynolds number. Therefore, it is more difficult to approximate the shedding mechanisms with a linear model. In addition to the harmonic decomposition, we also tried using Linear Stochastic Estimation and its variants (SLSE \cite{baars2014proper} and mtd-LSE \cite{durgesh2010multi}), but none gave a reliable estimate of the flow field.

\begin{figure}[h]
\centering
\includegraphics[width=\textwidth]{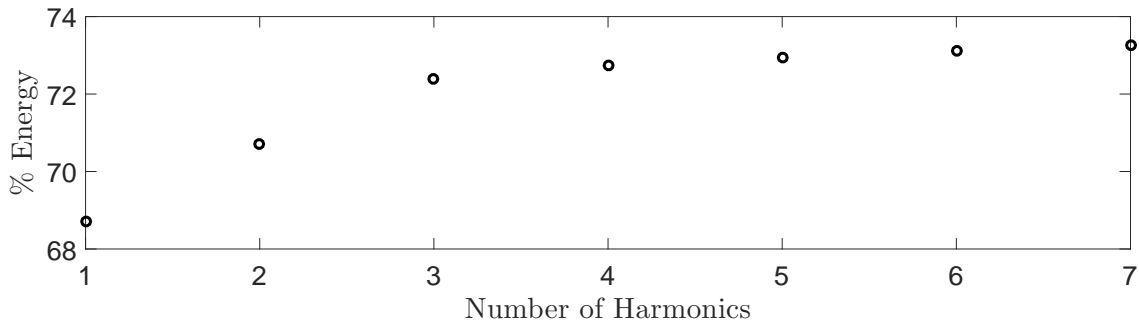}
\caption{Percentage of total energy captured with increasing number of harmonics involved for $Re=1036$.}
\label{fig:Re1000_energy}
\end{figure}

We first look at reconstruction using Harmonic Decomposition. Figure \ref{fig:Re1000_energy} shows the energy captured in the reconstructed flow as the number of harmonics retained is increased. A clear drop of the gradient can be found at the third harmonic, which means harmonics higher than the third harmonic contain significantly less energy. Therefore, only the first three harmonics are considered in the estimation, capturing $72.4\%$ of the total energy. The effect of the truncation order will be discussed in more detail later.

 
\begin{figure}[h]
\centering
\includegraphics[width=\textwidth]{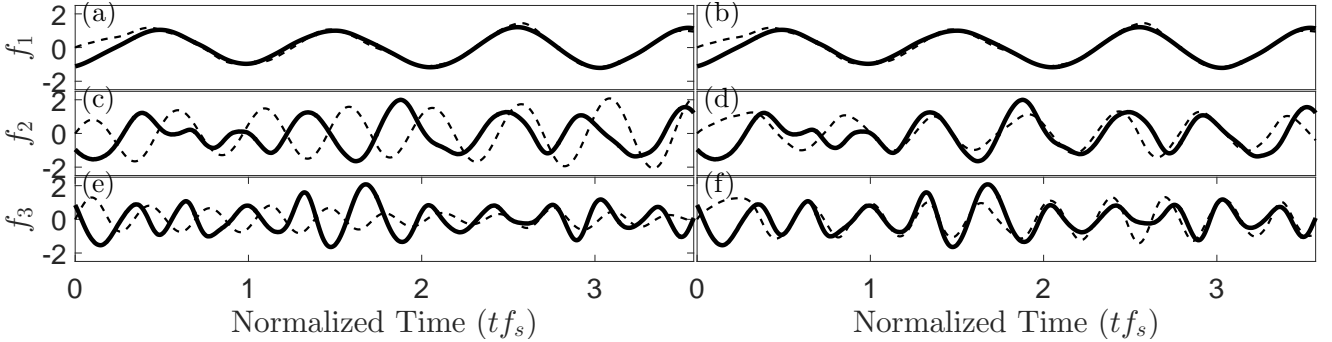}
\caption{Time dynamics of the Fundamental Harmonic(a,b), the second harmonic(c,d) and the third harmonic(e,f) from PIV(solid lines), Linear Estimation(a,c,e dashed lines) and Linear-Trigonometric Estimation(b,d,f dashed lines). Time has been normalized by the vortex shedding frequency ($t^*=t f_s$).}
\label{fig:Re1000harmonics}
\end{figure}

We first compare the estimate of the time dynamics to harmonic decomposition (Equation \ref{eq:xt_re100}), as shown in Figure \ref{fig:Re1000harmonics}. One observation is that the time dynamics computed directly from PIV measurement (thick solid lines) show both amplitude variation and phase shifting, which does not occur at $Re=100$. This indicates a disturbance to the linear model. Such disturbance could be the effect of turbulent structures. The other observation is that with good performance at the fundamental harmonic, LE fails at higher harmonics, whereas LTE still estimates the time dynamics at higher harmonics reasonably well. It again suggests strong nonlinear connections between the time dynamics of the fundamental harmonic and higher harmonics.

\begin{figure}[h]
\centering
\includegraphics[width=\textwidth]{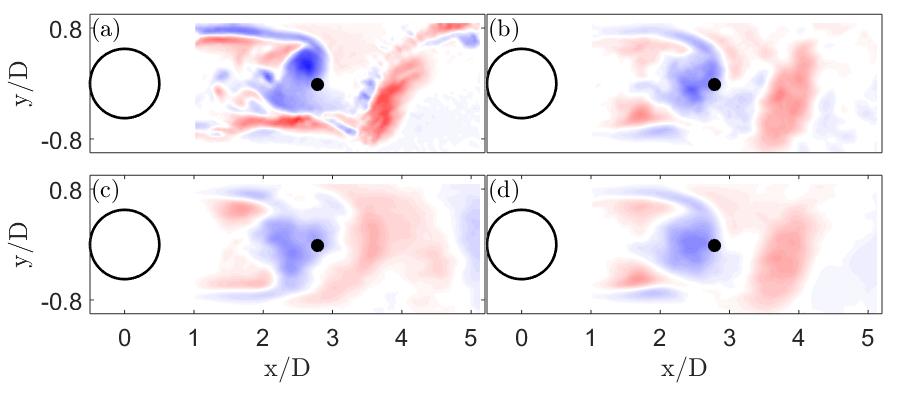}
\caption{The instantaneous vorticity fields from PIV (a), harmonic decomposition using the first three harmonics (b) and the estimates from Linear Estimation (c) and Linear-Trigonometric Estimation (d). Contour levels for all plots are $-0.26$ (blue) to $0$ (white) to $0.26$ (red).}
\label{fig:Re1000_IE_contourf}
\end{figure}

Now let us consider the instantaneous flow fields in physical space. Figure \ref{fig:Re1000_IE_contourf} shows instantaneous vorticity fluctuation fields from PIV, Harmonic Decomposition (3 harmonics), LE and LTE. It can be observed that Harmonic Decomposition (Figure \ref{fig:Re1000_IE_contourf}(b)) captures the large-scale vortex street reasonably well, but small turbulent structures are not captured. This explains why the first three harmonics only capture $72.4\%$ of the energy for this Reynolds number whereas it captured $99.9\%$ in the $Re=100$ case. The position of the large vortical structures are estimated well by both LE (Figure \ref{fig:Re1000_IE_contourf}(c)) and LTE (Figure \ref{fig:Re1000_IE_contourf}(d)), but LTE better estimates both the overall shape and inclination angle of the structures. This is consistent with the observation in the Re 100 case. Furthermore, even though discrepancies exist compared to the PIV field, the LTE estimate agrees well with the field from Harmonic Decomposition. This suggests that the estimation error is mostly contributed by the system modelling procedure rather than by the estimation of that model. Better performance might therefore be achieved with improved system modelling. 

\begin{figure}[h]
\centering
\includegraphics[width=\textwidth]{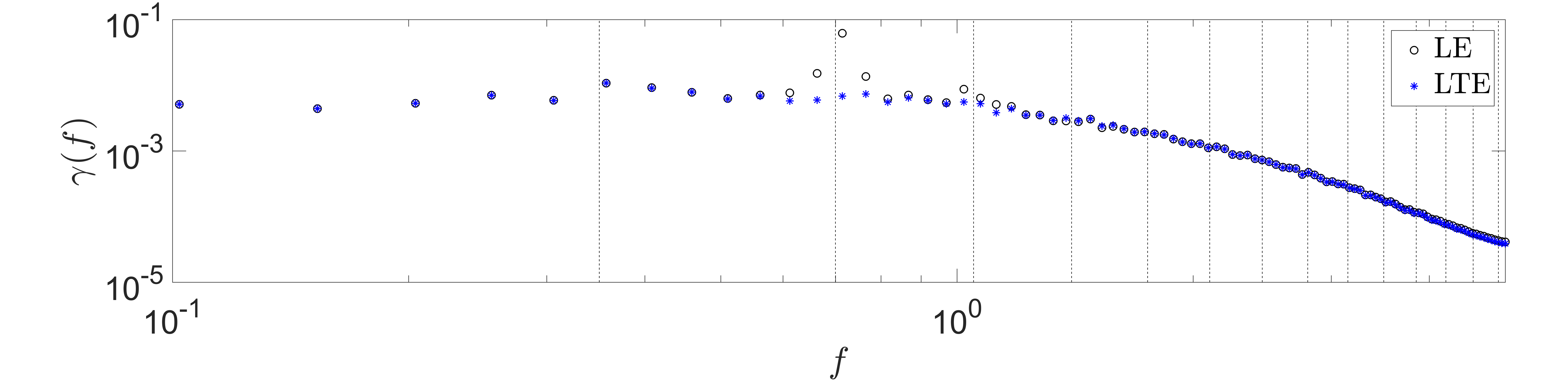}
\caption{Spectrum of the integrated estimation error from LE and LTE. All harmonic frequencies are marked by dashed lines.}
\label{fig:psd_error}
\end{figure}

A quantitative comparison of the estimator performance is shown in Figure \ref{fig:psd_error}. The lower $\gamma$ values at higher harmonics for LTE indicates better performance than LE at these harmonics, which is consistent with our previous observations.

\subsection{Effect of the Truncation Order}

\begin{figure}[h]
\centering
\includegraphics[width=.9\textwidth]{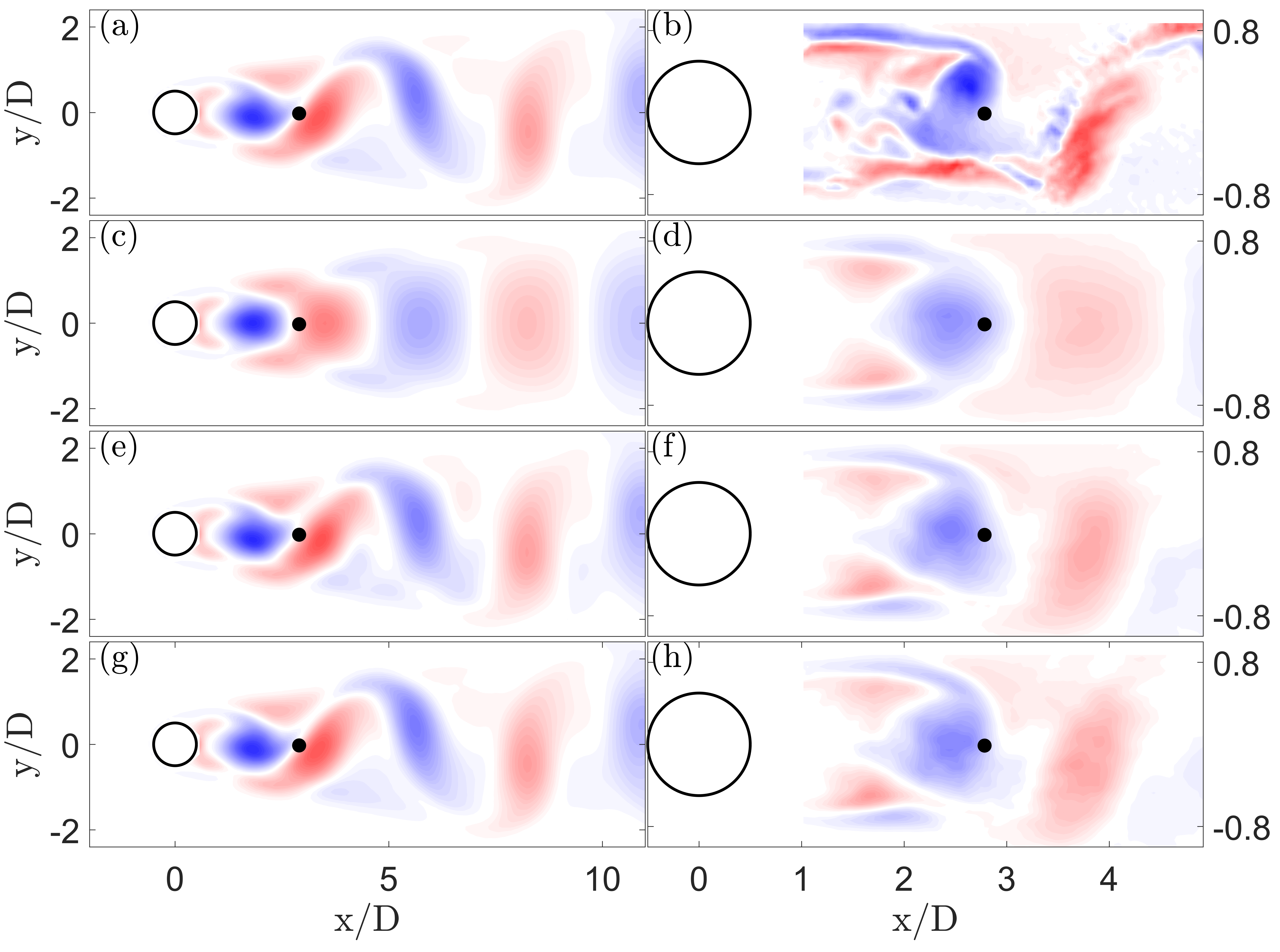}
\caption{The instantaneous vorticity fields at $Re=100$ (a,c,e,g) and at $Re=1036$ (b,d,f,h). DNS and PIV fields (a,b) are compared with the estimates of the fields using the fundamental harmonic alone (c,d), using the first three harmonics (e,f) and using the first seven harmonics (g,h). Contour levels are $-3$ (blue) to $0$ (white) to $3$ (red) for the $Re=100$ case (a,c,e,g), and $-0.26$ (blue) to $0$ (white) to $0.26$ (red) for the $Re=1036$ case (b,d,f,h). }
\label{fig:field_f1f2}
\end{figure}

Now we look at the influence of the number of harmonics on the estimation performance. Figure \ref{fig:field_f1f2} compares the estimate from LTE using different numbers of harmonics in both the $Re=100$ case and the $Re=1036$ case. Figure \ref{fig:field_f1f2}c,d show the estimate of a instantaneous vorticity field using the fundamental harmonic alone in the two cases. The flow patterns are symmetric with respect to the centre line and only the convective motion of the vortical structures is captured. The vorticity fields reconstructed using the first three harmonics, shown in Figure \ref{fig:field_f1f2}e,f, are able to capture the asymmtric flow features, such as vortex tilting and stretching. This indicates that higher harmonics should be considered if one intends to estimate cylinder wakes with correct asymmetric shedding patterns. As discussed in Section \ref{RE100} and Section \ref{RE1000}, the first three harmonics are sufficient to reconstruct the shedding patterns reasonably well. Including more harmonics may only improve the estimation accuracy slightly if at all. Figure \ref{fig:field_f1f2}g,h together demonstrate that using the first 7 harmonics has an insignificant effect on the estimator performance.

\subsection{Nonlinear interaction among harmonics}\label{s54}
In both the $Re=100$ case and the $Re=1036$ case, LTE provide reasonable estimates of the higher harmonics using only the time dynamics of the fundamental harmonic. This indicates the higher harmonics are in some sense slave to the fundamental harmonic. This is consistent with the findings in a recent paper by \cite{rosenberg2019role}. They found that the mode at the second harmonic can be generated by forcing a Navier-Stokes-based resolvent model with the triadic interactions generated by the fundamental harmonic. The forced resolvent mode agrees with the DMD mode reasonably well at the same frequency. From a Navier-Stokes-based point of view, their results support our suggestion of higher harmonics being "slave" to the fundamental harmonic. Another recent attempt of utilizing the nonlinear interaction among harmonics is by Meliga \cite{meliga2017harmonics}, who predicts the growth rates and dominant frequencies of cavity flows by considering the nonlinear interaction among harmonics as a feedback (via formation of Reynolds stresses) to a self-consistent model. 

\section{Conclusions}\label{s6}
Single-sensor based estimation of the cylinder wake has been considered for two cases: in simulations at $Re= 100$ and in experiments at $Re= 1036$. The estimation is based on a harmonic-based model, which is first used to demonstrate the effects of the higher harmonics on the vortex street patterns. The model is then used in Linear Estimation, which estimates the dynamics of the fundamental harmonic reasonably well in both cases. By further including the nonlinear relationship between harmonics, Linear-Trigonometric Estimation estimates the dynamics of the higher harmonics based on the fundamental harmonic, and then reconstructs the flow evolution with reasonable accuracy in both cases. By examining the overall estimation error in the frequency domain, it is found that the improvement of LTE over LE occurs at the higher harmonics. The fact that LTE yields better performance---particularly for the higher harmonics---indicates that, for the flow under consideration, the dynamics of the higher harmonics are in some sense slave to the dynamics of the fundamental harmonic.

We expect reasonable performance for Linear-Trigonometric Estimation for cylinder wakes with Reynolds numbers higher than $1036$. This is because we expect LTE to perform well when the flow contains energetic structures that are periodic in time and coherent in space. Although small-scale turbulent structures become increasingly important in cylinder wakes as Reynolds number increases, it is also known that periodic structures are still significant and energetically important even at $Re = 1.4\times 10^5$ \cite{braza2006turbulence}. Not only in cylinder wakes, LTE also has the potential to be applied in other types of flows in which the dominant flow structure is periodic, such as cavity flows.


\bibliographystyle{apsrev4-2}
\bibliography{bb}
\end{document}